\documentclass[12pt]{elsarticle} 
\usepackage{epsfig}
\usepackage{graphicx}
\usepackage{color}

\begin{document}

\title{Strength of correlations in pnictides and its assessment by theoretical calculations and  spectroscopy experiments}

\author {V.~I.~Anisimov$^1$, E.~Z.~Kurmaev$^1$, A.~Moewes$^2$, I.~A.~Izyumov$^1$}

\address{$^1$Institute of Metal Physics, Russian Academy of Sciences Ural Division,
620041 Yekaterinburg, Russia\\
$^2$Department of Physics and Engineering Physics, University of 
Saskatchewan, 116 Science Place Saskatoon,
Saskatchewan S7N 5E2, Canada
\\
}

\begin {abstract}

LDA+DMFT (Local Density Approximation combined with Dynamical
Mean-Field Theory) computation scheme  has been
used to calculate spectral properties of LaFeAsO -- the parent compound
of the new high-T$_c$ iron oxypnictides. The average Coulomb repulsion $\bar{U}=$3$\div$4~eV and Hund's
exchange $J$=0.8~eV  parameters for iron $3d$ electrons  were calculated using the \textit {first
principles} constrained density functional theory scheme in the Wannier
functions formalism.  DMFT calculations
using these parameters 
result in moderately correlated electronic structure with effective electron mass enhancement $m^*\approx$2 that is in
agreement with the experimental X-ray and photoemission spectra. Conclusion of moderate correlations strength is confirmed by the observation that pnictides experimental spectra agree well with corresponding spectra for metallic iron while being very different with Mott insulator FeO spectra.

\end {abstract}
\maketitle

Recent discovery of high-$T_c$ superconductivity   in   iron oxypnictides LaO$_{1-x}$F$_x$FeAs ~\cite {Kamihara-08} has stimulated
an intense experimental and theoretical activity. In a striking similarity with the high-$T_c$ cuprates,
the undoped LaFeAsO is not superconducting, but exhibits an antiferromagnetic
commensurate spin-density wave below 150~K~\cite {neutrons}. 
Only when electrons (or holes) are added to the system via doping,
antiferromagnetism is suppressed and superconductivity appears. As it is
generally accepted that the Coulomb correlations between the copper $3d$ electrons
are responsible for the anomalous properties of cuprates, it is tempting to
suggest that the same is true for the iron $3d$ electrons in LaFeAsO.

\begin {figure}
\includegraphics [width=0.99\textwidth]{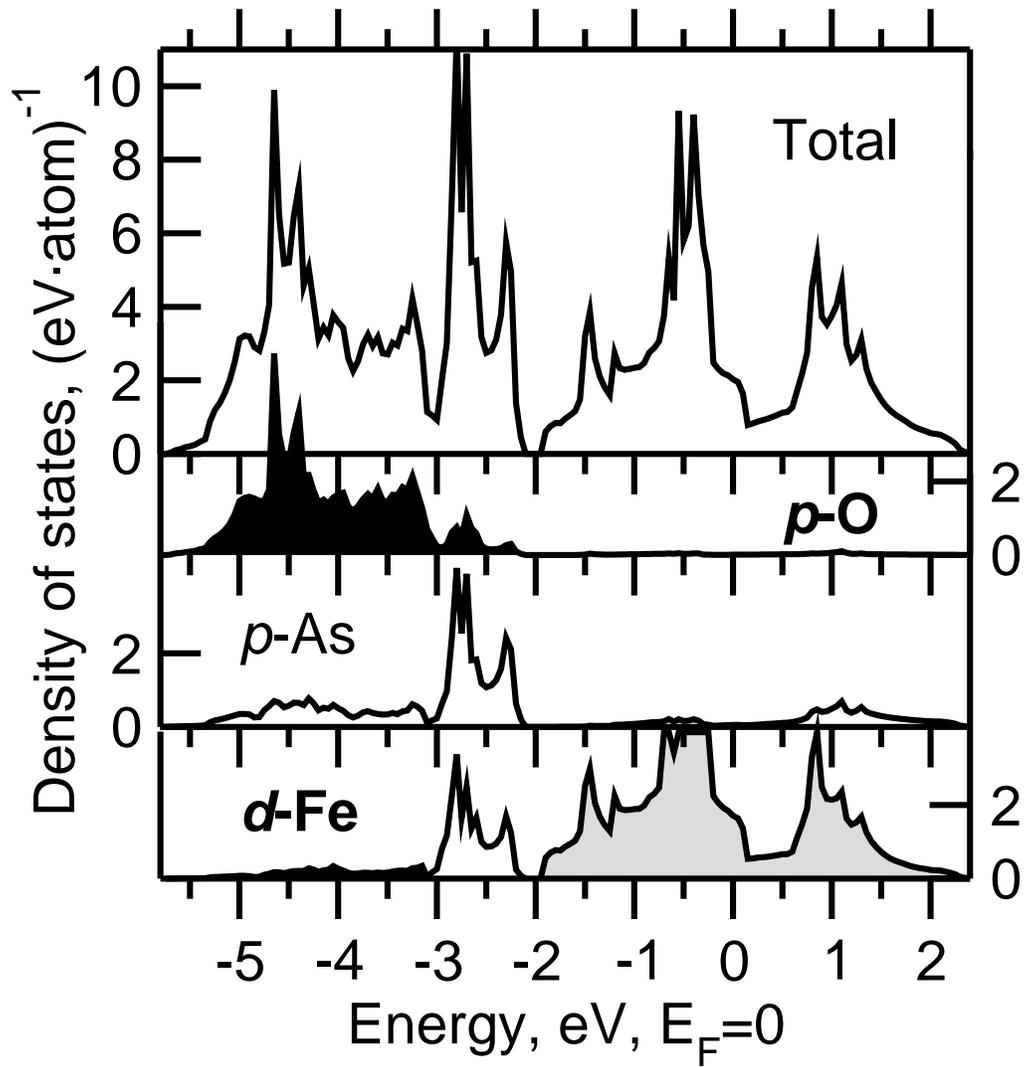}
\caption {Total and partial densities of states for LaFeAsO obtained in DFT
calculation in frame of LMTO method.}
\label {fig1} 
\end {figure}

The ratio of the Coulomb interaction $U$ and the band width $W$
determines the correlation strength. For $U/W<1$  
the system is weakly correlated and the results of Density Functional
Theory (DFT) calculations are enough to explain its electronic and
magnetic properties. However, if the $U$ value is comparable
with $W$ or even larger the system is in an intermediate or
a strongly correlated regime and the Coulomb interactions must be 
treated explicitly in the electronic structure calculations. 
The partially filled bands formed by Fe-$3d$ states in LaFeAsO
have a width of $\approx$4~eV (see the shaded area in the lower panel of
Fig.~\ref {fig1}), so the interaction parameter $U$ should be compared
with this value.

In practice, $U$ is often considered a free parameter to
achieve the best agreement between the calculated and measured properties of
the investigated system. However, the most attractive approach is to
determine the Coulomb interaction parameter $U$ from the \textit {first principles}. 
Two methods are generally used for this purpose: constrained DFT 
scheme~\cite {U-calc,anigun}, where the $d$-orbital
occupancies in DFT calculations are fixed to the certain values and $U$ is 
determined as a numerical derivative of the $d$-orbital energy over its occupancy, and 
Random Phase Approximation (RPA)~\cite {RPA}, where the screened Coulomb interaction
between the $d$-electrons is calculated using a perturbation theory. Recently, RPA calculations of the 
interaction parameter $U$ in LaFeAsO were reported~\cite {Arita}, estimating putting the $U$
value in the range 1.8$\div$2.7 eV. In 
Ref.~\cite {Haule08} it was proposed to use the $U$ of 4~eV
obtained in RPA calculations for metallic iron~\cite {Ferdi}. 
This value of the Coulomb parameter (with Hund's exchange parameter
$J$=0.7~eV) was used in Dynamical Mean-Field Theory (DMFT)~\cite {DMFT}
calculations for LaFeAsO \cite {Haule08, Craco08, DMFT-our}. 
These studies find the iron 3$d$ electrons to be in an intermediate or a strongly
correlated regime, as can be expected for the Coulomb parameter
$U$=4~eV and the Fe-$3d$ band width of $\approx$4~eV. 

To estimate the correlation strength one can compare experimental spectra with the densities
of states (DOS) obtained in DFT calculations. For strongly correlated
materials additional features in the experimental photoemission, X-ray absorption and
optical spectra appear that are absent in the DFT DOS. These features are interpreted 
as lower and upper Hubbard bands. If no Hubbard bands are observed and the DOS obtained in 
DFT calculations describes the experimental spectra satisfactorily the material is 
considered to be in a weakly correlated regime. 
LaFeAsO was studied by soft X-ray absorption and emission
spectroscopy~\cite {X-ray}, X-ray absorption (O $K$-edge)
spectroscopy~\cite {exp-U}, and photoemission spectroscopy~\cite
{Malaeb-08}. All these studies conclude that DOS obtained in DFT
calculations agrees well with the experimental spectra and the
estimated value of the Coulomb parameter is less than 1~eV~\cite {exp-U}. Such
a contradiction with the DMFT results~\cite {Haule08, Craco08,DMFT-our} using $U$=4~eV shows that the \textit {first principles} calculation of 
the Coulomb parameter $U$ for LaFeAsO is needed to assess
the strength of correlations in this material. Results of such
calculations using the constrained DFT methods are reported in the present
work. 

A source of uncertainty in the constrained DFT scheme is the
definition of atomic orbitals whose occupancies are fixed and energies are
calculated. In some DFT methods, like Linearized Muffin-Tin Orbitals
(LMTO), these orbitals could be identified with LMTOs. However, for other 
basis sets for example plane waves in the
pseudopotential methods one should use a more general definition of the localized
atomic-like orbitals such as Wannier functions~\cite {Wannier37} (WFs). A
practical way to calculate WFs for specific materials using projection of
atomic orbitals on the Bloch functions was developed in Ref.~\cite
{MarzariVanderbilt}. 

In Fig.~\ref {fig1} the total and partial DOS for LaFeAsO obtained in  LMTO
basis are shown. The crystal field splitting of the Fe-$3d$ states in
this material is rather weak ($\Delta_{cf}$=0.25~eV) and all five $d$
orbitals of iron form a common band in the energy range ($-$2, $+$2)~eV
relative to the Fermi level (see the gray region in the bottom panel of
Fig.~\ref {fig1}). There is a strong hybridization of the iron $t_{2g}$
orbitals with the $p$ orbitals of arsenic, the effect of which
becomes apparent in the energy
interval ($-$3, $-$2)~eV (the white region in the bottom panel of Fig.~\ref
{fig1}) where the As-$p$ band is situated. A weaker
hybridization with the oxygen $p$ states can be seen in ($-$5.5, $-$3)~eV energy
window (the black region in the bottom panel of Fig.~\ref {fig1}).

We have calculated the average Coulomb interaction $\bar{U}$ and Hund's exchange $J$
parameters in WFs basis using the constrained DFT procedure with fixed
occupancies of WFs of $d$ symmetry. For this purpose we have used two
computational schemes based on two different DFT methods. The first
involves linearized muffin-tin orbitals produced by the TB-LMTO-ASA
code~\cite {LMTO}; the corresponding WFs calculation procedure is described in
details in Ref.~\cite{WF-LMTO}. The second, based on the pseudopotential plane-wave method PWSCF, as
implemented in the Quantum ESPRESSO package~\cite {PW}, is described in
Ref.~\cite{WF-PW}. The difference between results obtained with the
two schemes gives an estimate of the uncertainty of $\bar{U}$ and $J$
determination.

The WFs are defined by the choice of Bloch functions Hilbert space and by a
set of trial localized orbitals that will be projected onto these Bloch
functions ~\cite {WF-LMTO}. We have performed calculations for  all bands in energy window
($-$5.5, $+$2)~eV that are formed by O-$2p$, As-$4p$ and Fe-$3d$ states and
correspondingly full set of O-$2p$, As-$4p$ and Fe-$3d$ atomic orbitals to
be projected on the Bloch functions for these bands. This corresponds to
an extended model where in addition to the Fe-$d$ orbitals all the $p$-orbitals are
included as well.

Constrain DFT calculations using TB-LMTO-ASA method gave values
$\bar{U}$=3.10, $J$=0.81 while calculations with PWSCF gave  $\bar{U}$=4.00, $J$=1.02. The difference between those two sets of parameter values shows the "error bar" in definition of  Coulomb interaction parameters. In the following we have used results obtained in TB-LMTO-ASA calculations.

In the constrained DFT calculations an average Coulomb interaction
$\bar{U}$ is obtained which can be approximated\cite{anigun} as $\bar{U}=F^0-J/2$, where $F^0$ is 0th Slater integral. Hence $F^0$ can be calculated as $F^0=\bar{U}+J/2$ and this gives $F^0$=3.5~eV
and $J$=0.81~eV. With this set of parameters Coulomb interaction matrix $U_{mm'}$ was calculated and used in 
LDA+DMFT~\cite {LDA+DMFT} calculations (for a detailed description of the
present computation scheme see Ref.~\cite{WF-LMTO}). The DFT band
structure was calculated within the TB-LMTO-ASA method~\cite {LMTO}.
Crystal structure parameters of Ref.~\cite{Kamihara-08} were used.

A double-counting term $\bar{U}(n_{DMFT}-\frac{1}{2})$ was used to obtain the noninteracting Hamiltonian \cite{WF-PW}.
Here $n_{DMFT}$ is  the total number of $d$-electrons obtained
selfconsistently within the LDA+DMFT scheme. The effective impurity model
within the DMFT was solved by Hirsch-Fye QMC method~\cite
{HF86}.
In the present implementation of the QMC impurity solver the Coulomb
interaction between different orbitals on the same atom is limited to
density-density terms, i.e. the form $\sum_{m\sigma,m'\sigma '}U^{\sigma\sigma '}_{mm'}(F^0,J)\hat{n}_{m\sigma}\hat{n}_{m'\sigma '}$.
In particular, this means that the coupling between the local spins is of Ising and not
Heisenberg type. Since this is a significant approximation a few comments are in order
especially concerning the question whether this approximation underestimates or overestimates
the many-body renormalization of quasi-particle bands. We argue quite generally
that introducing the spin-flip exchange (and other interaction terms beyond density-density)
allows electrons to avoid each other more efficiently and thus the electron propagation
through the crystal is inhibited less than with Ising exchange. As an example one can imagine
two electrons with opposite spin in different orbitals. Allowing the spin-flip exchange
the Ising-only interaction energy $U'$ can be reduced to $U'-J$ by forming the 
triplet state with $S_z=0$. This argument is also supported by recent numerical studies of two-band model
comparing full Coulomb and Ising-only interaction terms~\cite{Ising}.
Therefore we expect our results rather to overestimate than underestimate the quasi-particle 
renormalization.

Calculations  were performed at the inverse
temperature $\beta$=10~$eV^{-1}$. The interval
$0<\tau<\beta$ was divided into 100 slices.  $6 \cdot 10^6$ QMC sweeps were
used in self-con\-sis\-ten\-cy loop within the LDA+DMFT scheme and $12
\cdot 10^6$ of QMC sweeps were used to calculate the spectral functions.

\begin {figure}
\includegraphics [width=0.8\textwidth]{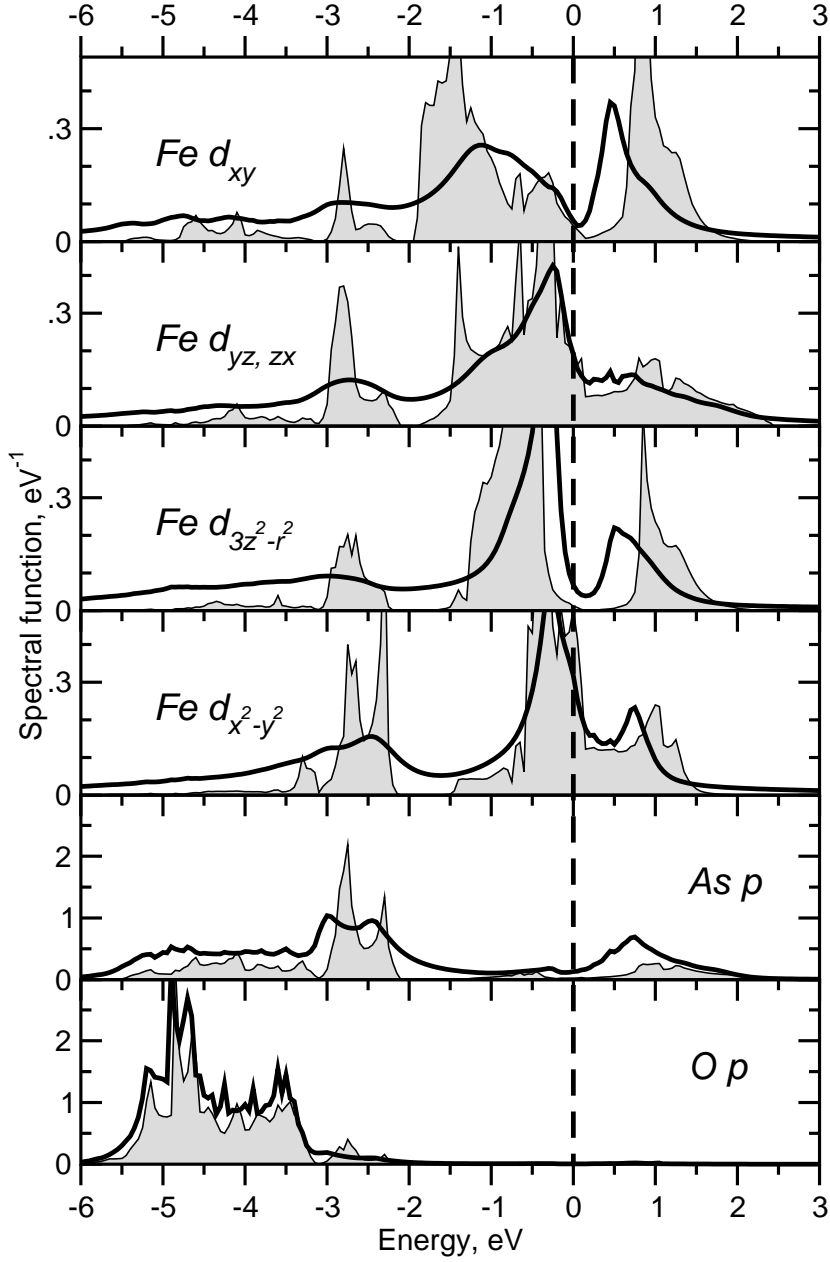}
\caption {Partial densities of states for Fe-$3d$ , As-$4p$ and O-$2p$ states
obtained within the DFT (filled areas) and LDA+DMFT orbitally resolved
spectral functions  (bold lines).}
\label {LDA+DMFT-dos_p-d} 
\end {figure}

The results of the  LDA+DMFT calculations
are presented in Fig.~\ref{LDA+DMFT-dos_p-d}. The effect of correlations on the electronic
structure of LaFeAsO is minimal: there are relatively small changes of peak positions
for $3z^2-r^2$, $xy$ and $x^2-y^2$ orbitals (a shift toward the Fermi
energy) and practically unchanged spectral functions 
for the $yz, zx$ bands. There is no appearance of either Kondo resonance peak
on the Fermi level or Hubbard bands in the spectrum, the features in Fe-$d$ spectral functions below -2 eV correspond to hybridization with  As-$p$ and O-$p$ bands. 
The reason for such weak correlation effects in spite of the relatively strong Coulomb interaction 
is very a strong hybridization of the Fe-$d$ orbitals with As-$p$ states 
(see the peaks in the Fe-$d$ spectral function in the -2$\div$-3 eV range corresponding to mixing with As-$p$ bands). 
The hybridization provides an additional very efficient channel for screening of the Coulomb interaction between  Fe-$d$ electrons.

These observations agree with the results of soft X-ray absorption and emission
spectroscopy study~\cite {X-ray}. It was concluded there that LaFeAsO does
not represent a strongly correlated system since the Fe $L_3$ X-ray emission
spectra do not show any features that would indicate presence of the
lower Hubbard band or a sharp quasiparticle peak that were predicted by the
LDA+DMFT analysis~\cite {Haule08, Craco08, DMFT-our}. The comparison of the X-ray absorption spectra (O $K$-edge) with
the LDA calculations gave~\cite {exp-U} an upper limit of the effective on-site
Hubbard $U\approx$1~eV. The photoemission study of LaFeAsO
suggests~\cite {Malaeb-08} that the line shapes of Fe $2p$ core-level
spectra correspond to an itinerant character of Fe $3d$ electrons. It was
demonstrated there that the valence-band spectra are generally consistent
with the band-structure calculations except for shifts of the Fe $3d$-derived
peaks towards the Fermi level. Such a shift is indeed observed in our LDA+DMFT spectra (Fig.~\ref
{LDA+DMFT-dos_p-d}).

The resonant inelastic  X-ray scattering (RIXS) measurements of the LaOFeAs ~\cite {X-ray}, LiFeAs and NaFeAs ~\cite {kur2}, CaFe$_2$As$_2$ ~\cite {kur3} were performed at the soft X-ray fluorescence endstation at Beamline 8.0.1 of the Advanced Light Source at Lawrence Berkeley National Laboratory ~\cite {kur4}. We have measured the resonant and non-resonant Fe $L_{2,3}$ ($3d4s  \rightarrow 2p_{1/2,3/2}$ transition) X-ray emission spectra (XES). The instrument resolution for Fe $L_{2,3}$ X-ray emission spectra was 0.8 eV. X-ray absorption spectra (XAS) were measured in the total fluorescence yield mode with a resolving power $E/\Delta E$=5000. 

\begin {figure}
\includegraphics [width=0.99\textwidth]{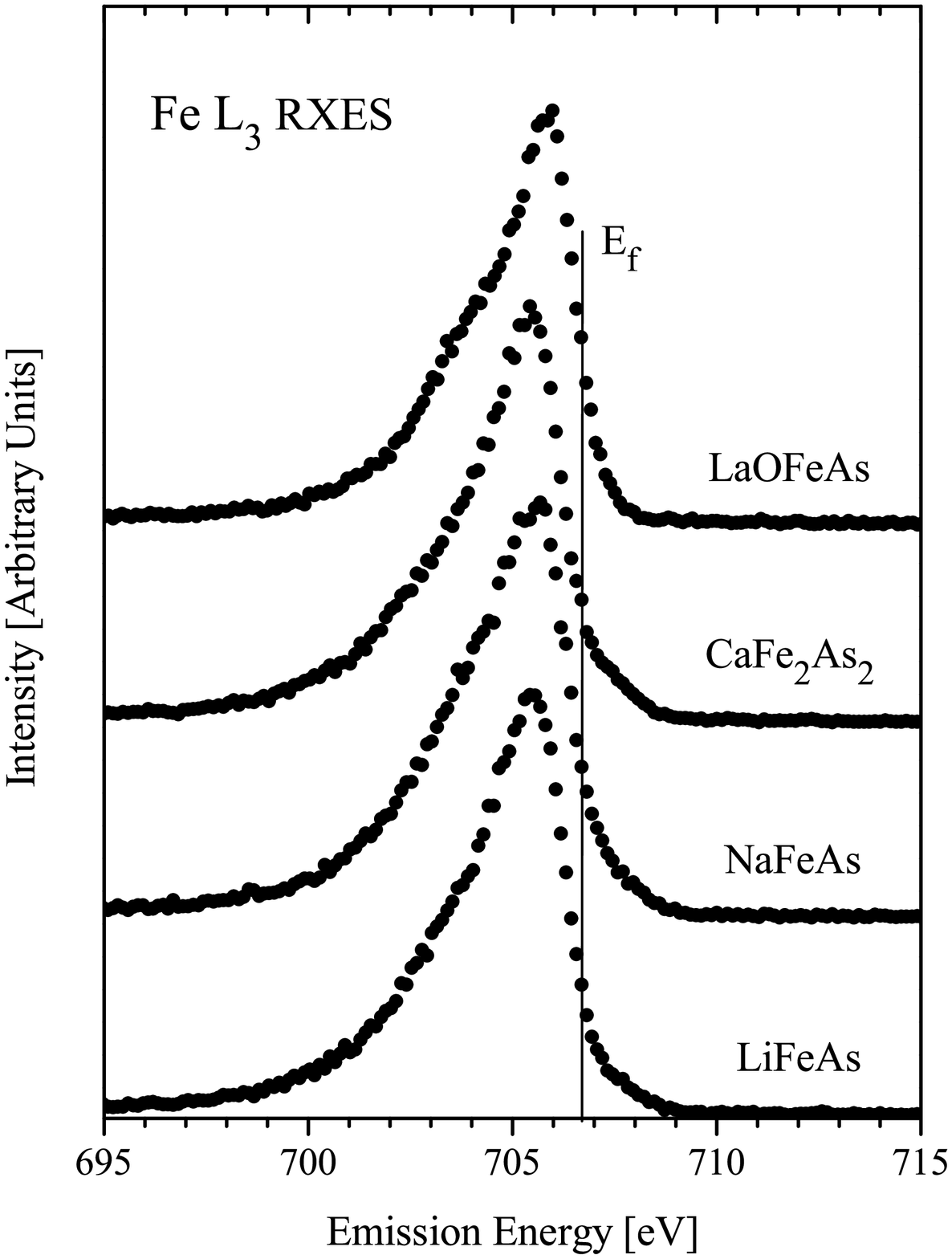}
\caption {Fe L$_3$ XES of LaOFeAs ~\cite {X-ray}, CaFe2As2 ~\cite {kur3}, LiFeAs and NaFeAs ~\cite {kur2}.}
\label {Fig1} 
\end {figure}

Resonantly excited Fe $L_{3}$ XES spectra of LaOFeAs, CaFe$_2$As$_2$, NaFeAs and LiFeAs which directly probe the distribution of occupied Fe 3$d$-states are presented in Fig.~\ref
{Fig1}. The position of the Fermi level on the spectral curves is determined using the XPS Fe 2$p$ binding energy for CaFe$_2$As$_2$ ($E_b$=706.7 eV) ~\cite {kur3}. One can see that the intensity maximum of  Fe $L_{3}$ XES is located within 0.9-1.25 eV with respect to the Fermi level which demonstrates that the Fe 3d-states dominate at the top of the valence band for both one-layered and two-layered FeAs-systems. For all investigated compounds Fe $L_{3}$ X-ray emission spectra do not show any features that would indicate the presence of the lower Hubbard band or a sharp quasiparticle peak.

\begin {figure}
\includegraphics [width=0.99\textwidth]{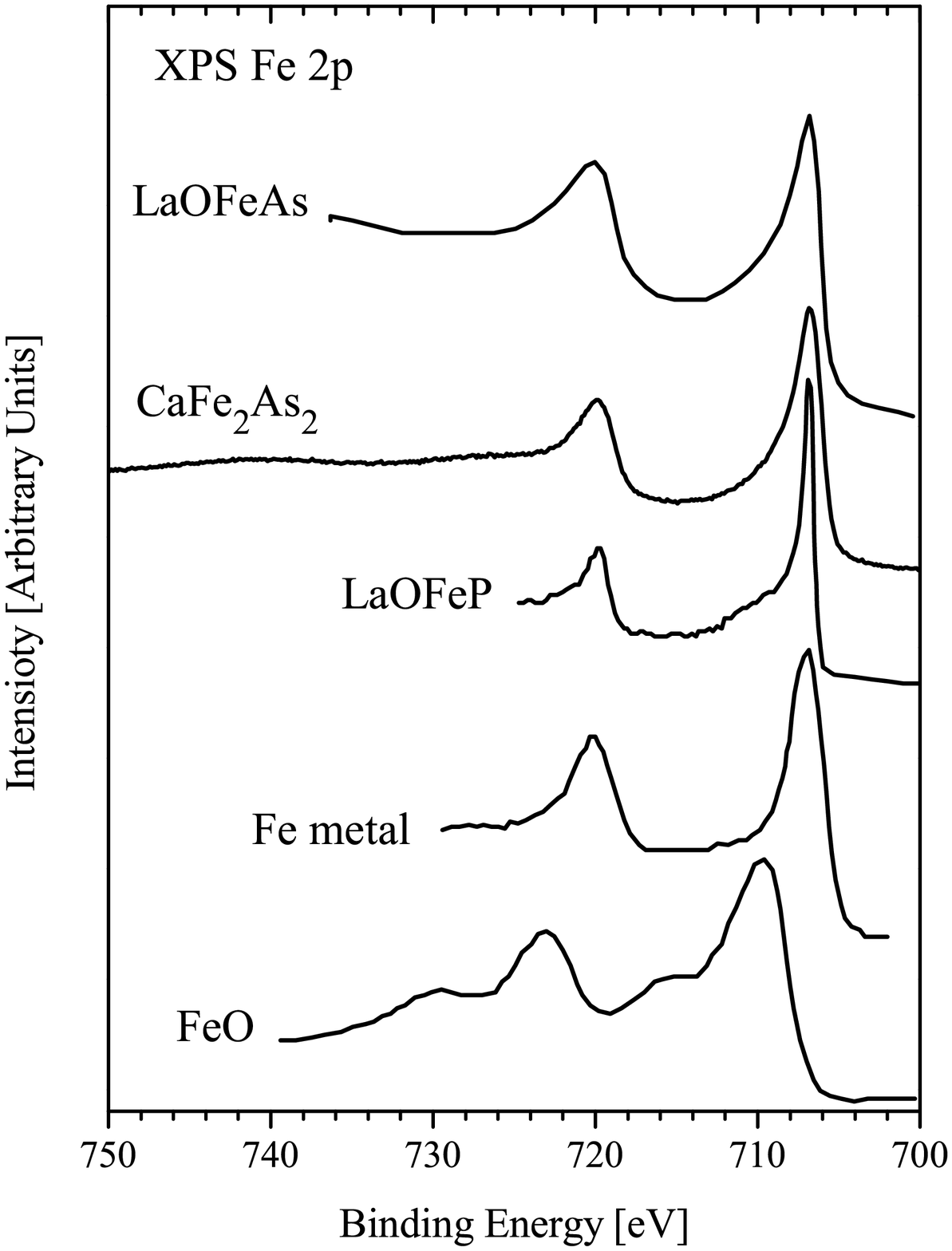}
\caption {XPS Fe $2p$ of LaOFeAs ~\cite {Malaeb-08}, CaFe2As2 ~\cite {kur3}, 
LaOFeP ~\cite {kur5}, FeO ~\cite {kur6} and Fe metal ~\cite {kur7}.}
\label {Fig2} 
\end {figure}

The XPS Fe 2$p$ core level spectra of LaOFeAs ~\cite {Malaeb-08}, CaFe$_2$As$_2$ ~\cite {kur3}, LaOFeP ~\cite {kur5} and FeO ~\cite {kur6} (see Fig.~\ref
{Fig2})  don't show any presence of satellites typical for correlated systems (such as FeO) and very similar to that of metallic Fe ~\cite {kur7}. Therefore the line shapes of XPS Fe 2$p$ core-level spectra of iron pnictides correspond to an itinerant character of Fe 3$d$ electrons. 

\begin {figure}
\includegraphics [width=0.99\textwidth]{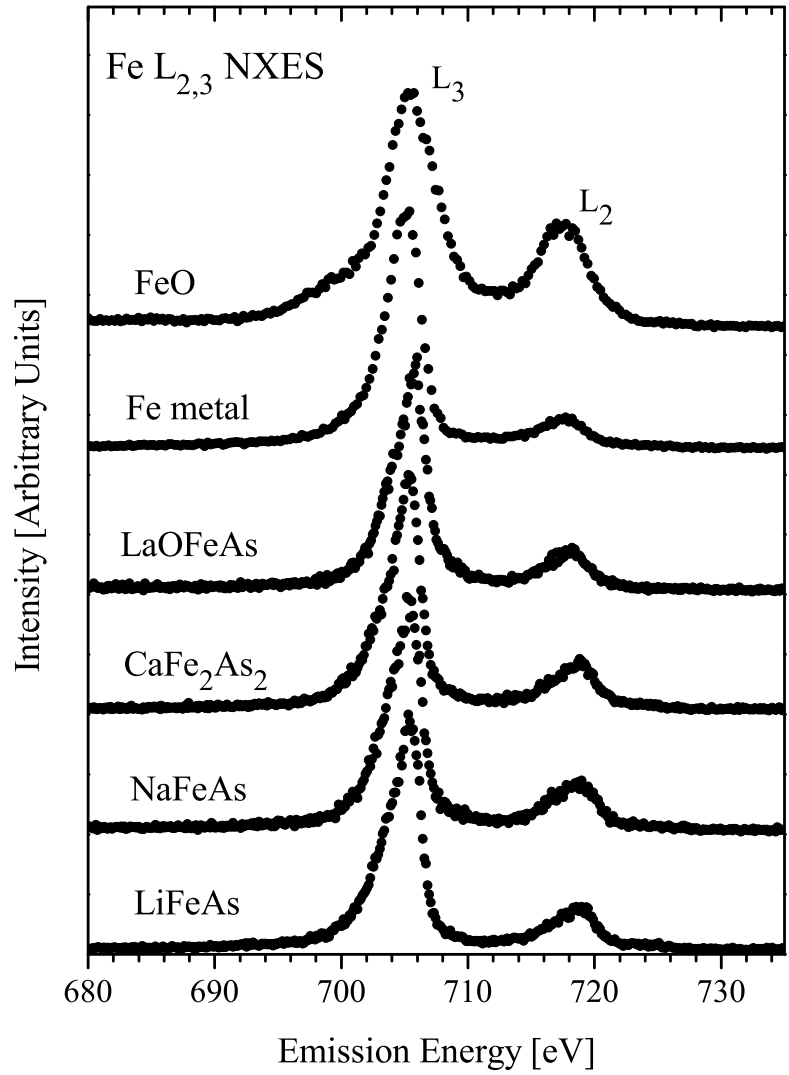}
\caption {Fe L$_{2,3}$ NXES of FeAs-systems and reference samples.}
\label {Fig3} 
\end {figure}

This conclusion is supported by comparison of non-resonant Fe $L_{2,3}$ XES of LaOFeAs, CaFe$_2$As2, NaFeAs and LiFeAs (Fig.~\ref
{Fig3}). The two main bands located around 705 and 718 eV correspond to Fe $L_3$ ($3d4s \rightarrow2p_{3/2}$ transition) and Fe $L_2$ ($3d4s \rightarrow 2p_{1/2}$ transition) normal emission lines, respectively, separated by spin-orbital splitting of Fe 2$p$. For free atoms the relative intensity ratio of $L_2$ and $L_3$ XES lines, I($L_2$)/I($L_3$), is determined only by the statistical population of $2p_{1/2}$ and $2p_{3/2}$ levels and therefore should be equal to 1/2. In metals the radiationless $L_2L_3M_{4,5}$ Coster-Kronig (C-K) transitions strongly reduce the intensity ratio I($L_2$)/I($L_3$) ~\cite {kur8}. One can see from the Fig. 3 that the intensity ratio I($L_2$)/I($L_3$) is almost identical for all above mentioned FeAs-compounds and more close to that of Fe metal than to FeO.

All available photoemission measurements of iron pnictides ~\cite {Malaeb-08,kur3,kur5,PES,kur10,kur11} as well as FeSe ~\cite {kur12} which probe a total density of states show a very similar fine structure which is consistent with LDA calculations. The Fe 3$d$ partial density of states of LaOFeAs determined by indirect way as a difference of photoemission spectra measured at and below Fe$3p \rightarrow3d$ resonance is found to be different in Refs. ~\cite {Malaeb-08} and ~\cite {kur11}. Authors ~\cite {kur11} have revealed in difference spectrum a broad peak centered at  7 eV and satellite at  12 eV which are attributed to the incoherent part of Fe 3$d$ states and contribution of super-Coster-Kronig Fe $3p-3d$ Auger transition (similar to that of FeO ~\cite {kur13}), respectively, which are considered as  an indication of relatively strong correlation effects in this material. On the other hand, in difference spectrum obtained in ~\cite {Malaeb-08} no broad peak at  7 eV is detected and spectral feature at  12 eV is  found to be consistent with calculated As $4s$ low energy subband.

\begin {figure}
\includegraphics [width=0.9\textwidth]{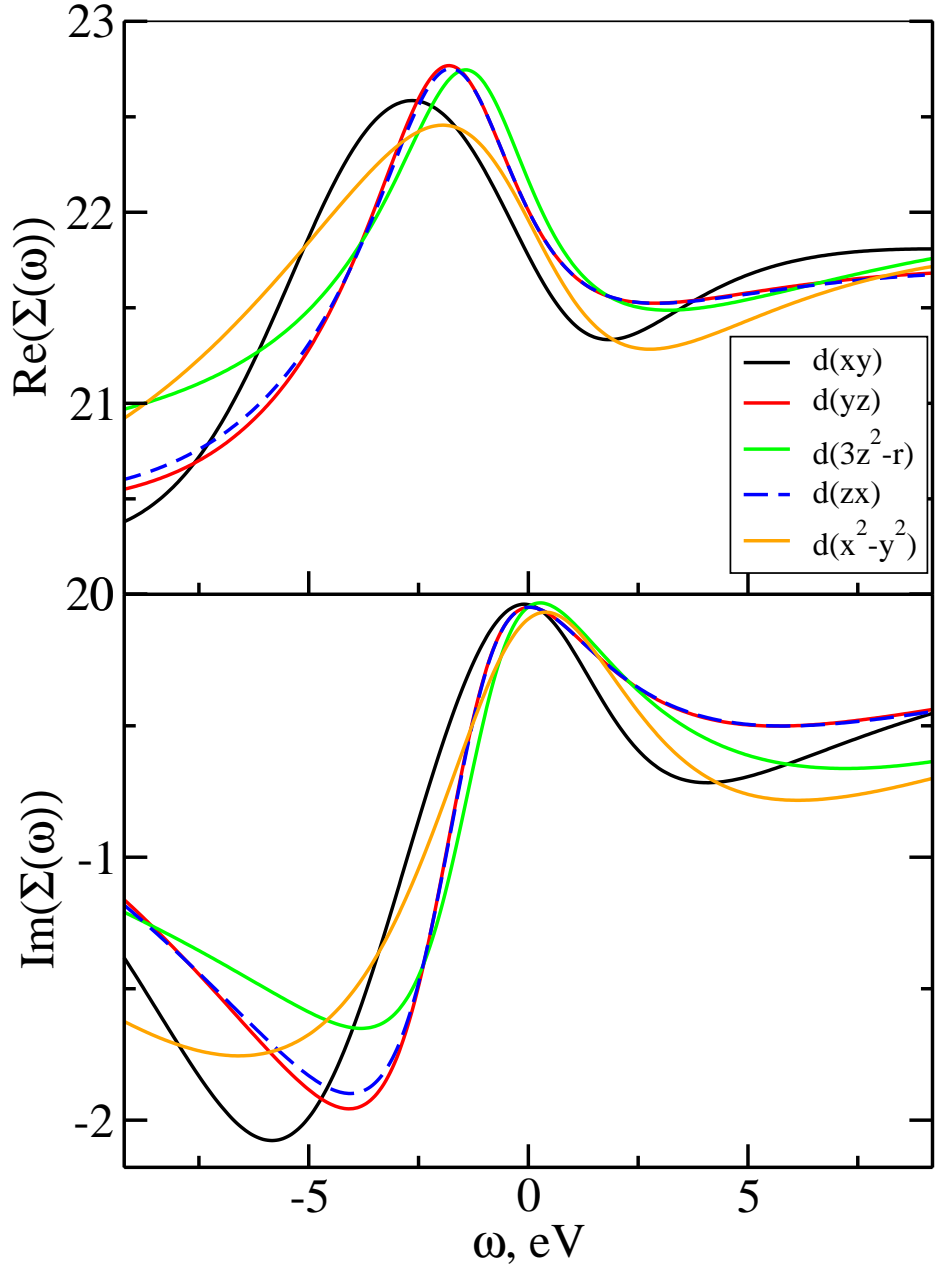}
\caption {(Colour online) Real (upper panel) and imaginary (lower panel)
parts of LDA+DMFT self energy interpolated on real axis with the use of
Pad\'e approximant .}
\label {p-d-sigma} 
\end {figure}

The behavior of the real part of self energy near zero frequency
$\Sigma(\omega)|_{\omega \rightarrow 0}$ provides an important
information about band narrowing and renormalization of the electron mass.
Pad\'e approximant~\cite {pade} was used to obtain the self
energy on the real frequency axis. The results  are presented in
Fig.~ \ref {p-d-sigma}. The calculated values of
the quasiparticle renormalization factor
$Z=(1-\frac{\partial\Sigma(\omega)}{\partial\omega}|_{\omega=0)})^{-1}$ are
found to be  0.56, 0.54,
0.45, 0.56  for $d_{xy}$, $d_{yz}$ (or $d_{zx}$),
$d_{3z^2-r^2}$, $d_{x^2-y^2}$ orbitals, respectively. These values  agree well with the effective narrowing of the
LDA+DMFT spectral functions relative to LDA DOS (Fig.~\ref {LDA+DMFT-dos_p-d}).
The effective mass enhancement $m^*=Z^{-1}$ are 1.78, 1.85, 2.22, 1.95
 for $d_{xy}$, $d_{yz}$ (or $d_{zx}$),
$d_{3z^2-r^2}$, $d_{x^2-y^2}$ orbitals, respectively,
agrees well with the mass enhancement factor between 1.7 and 2.1 reported in the dHvA study~\cite{dHvA} and also with the results of Angle-Resolved Photoemission Spectroscopy (ARPES) for pnictides ~\cite{kur11,kur5} where overall bandwidth was found to be reduced by a factor of 2.
 The $d_{x^2-y^2}$ orbital has the largest effective mass and
exhibits the most evident narrowing of LDA spectrum (see Fig.~\ref
{LDA+DMFT-dos_p-d}). This orbital has its lobes directed into the empty space
between nearest iron neighbors in the Fe plane. Hence it has the weakest
overlap, the smallest band width, and the largest $U/W$ ratio.

The small effective mass
enhancement shows that LaFeAsO is a moderately correlated
system in contrast to the results of LDA+DMFT calculation~\cite {Haule08}
by Haule {\it et al} where a strongly renormalized low energy band with a
fraction of the original width ($Z\approx$ 0.2-0.3) was found while most of
the spectral weight was transferred into a broad Hubbard band at the binding
energy $\approx$4 eV. Authors of Ref.~\cite{Haule08} report that
"slightly enhanced Coulomb repulsion ($U$= 4.5 eV) opens the gap" so that
the system is in strongly correlated regime on the edge of a metal-insulator
transition. 
It is difficult to understand why two LDA+DMFT calculations gave so different results because descriptions of calculation details in  Haule {\it et al} paper are very short. The only decisive way solve this problem would be third party independent calculation.

\begin {figure}
\vspace {5mm}
\includegraphics [width=0.99\textwidth]{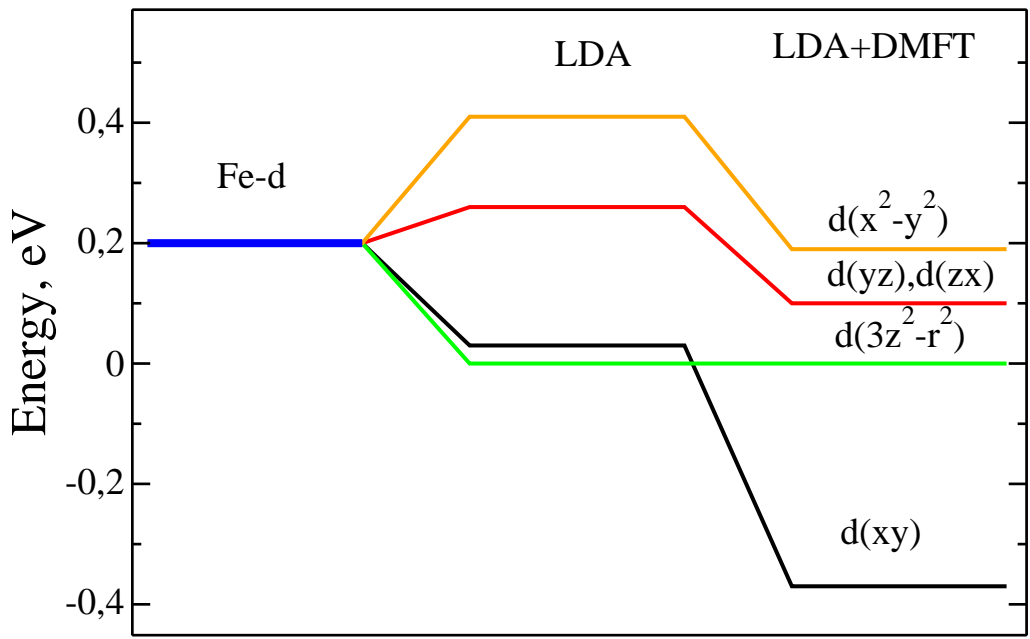}
\caption {(Colour online) Splitting of Fe-$d$ orbitals obtained in LDA and
LDA+DMFT .}
\label {split} 
\end {figure}

In LaFeAsO the iron ion is tetrahedrally coordinated with four As ions exhibiting
a slight tetragonal distortion. In the tetrahedral symmetry
group T$_d$ the five $d$-orbitals should be split by the crystal field
into a low-energy doublet of $3z^2-r^2$, $xy$ corresponding to
the $e_g$ irreducible representation and a high-energy triplet of
$x^2-y^2$, $xz$, $yz$ belonging to the $t_{2g}$ representation. We have calculated
the WF orbital energies and have found that the $t_{2g}$--$e_g$ crystal field
splitting is very small $\Delta_{cf}\approx$0.25~eV. The slight
tetragonal distortion of the tetrahedron leads to an additional splitting of
the $t_{2g}$ and $e_g$ levels with the following orbital energies
(the energy of the lowest $3z^2-r^2$ orbital is set to zero):
$\varepsilon_{3z^2-r^2}$=0.00~eV, $\varepsilon_{xy}$=0.03~eV,
$\varepsilon_{xz,yz}$=0.26~eV, $\varepsilon_{x^2-y^2}$=0.41~eV. The correlation
leads not only to narrowing of the bands but also to substantial shifts of the Fe-$d$ orbitals energies. 
Adding the $Re(\Sigma(0))$ to the LDA orbital
energies results in $\varepsilon_{3z^2-r^2}$=0.00~eV,
$\varepsilon_{xy}$=-0.37~eV, $\varepsilon_{xz,yz}$=0.10~eV,
$\varepsilon_{x^2-y^2}$=0.20~eV (see Fig.~\ref {split}). Note that the
actual band shifts are smaller due to the $p$-$d$ hybridization.

\begin {figure}
\vspace{5mm}
\includegraphics [width=0.8\textwidth]{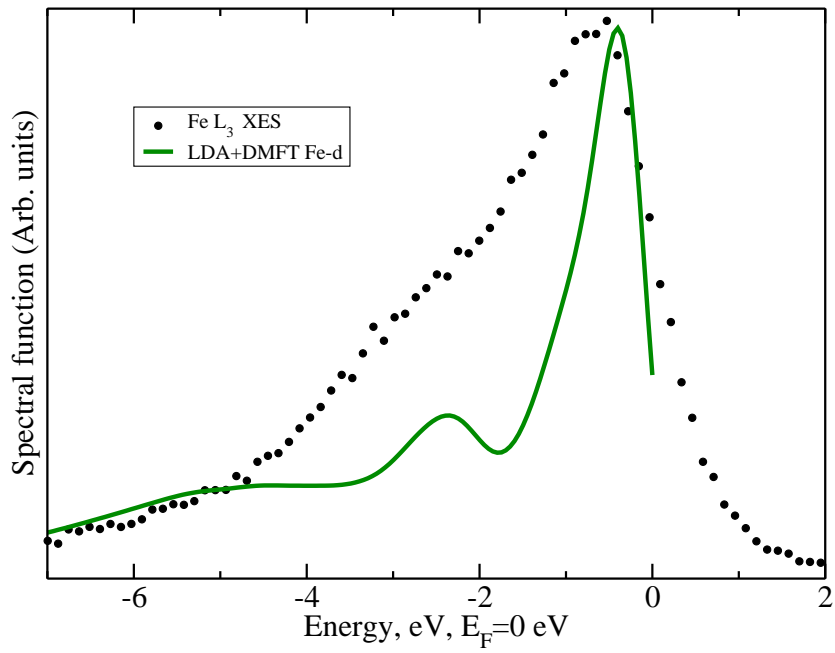}
\caption {(Colour online) Calculated Fe-$d$ LDA+DMFT  spectral function (solid line) and experimental Fe
L$_3$ XES spectrum (circles) from Ref.~\cite{X-ray}.}
\label {d-spec} 
\end {figure}

\begin {figure}
\includegraphics [width=0.8\textwidth]{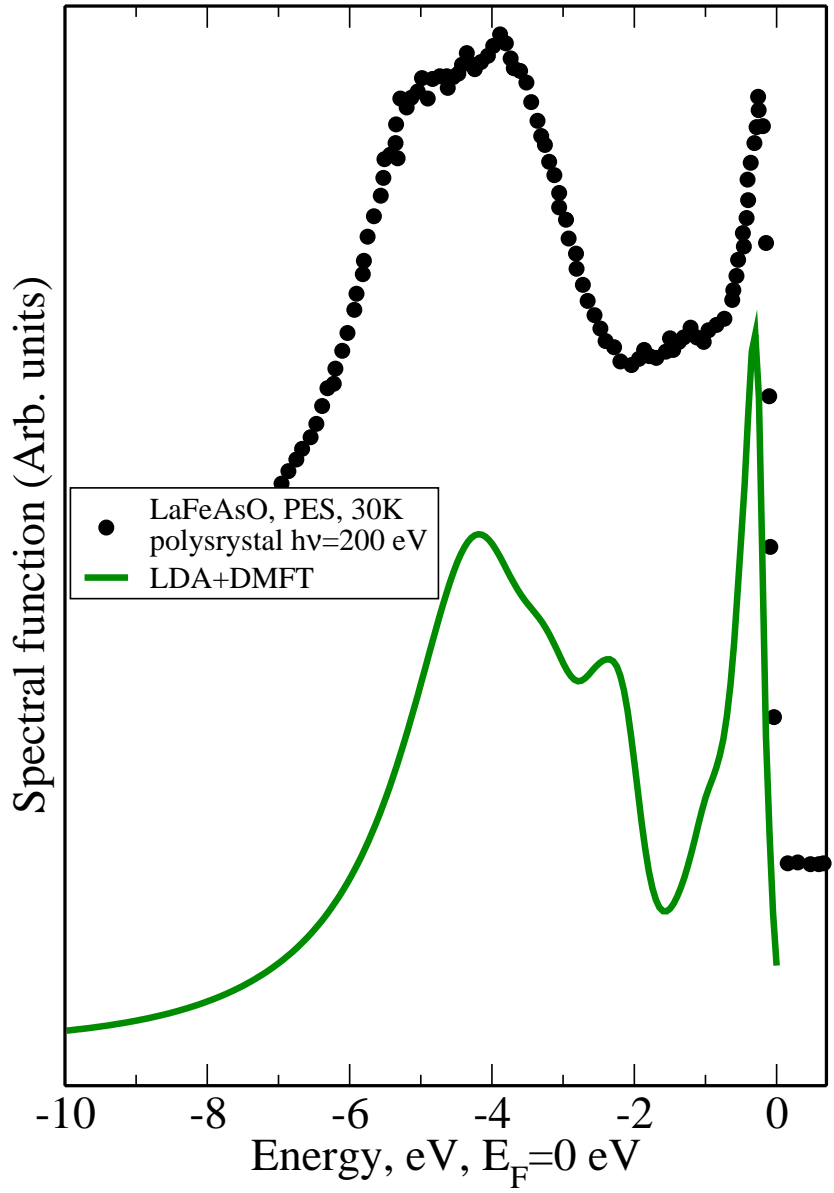}
\caption {(Colour online) Calculated total LDA+DMFT  spectral function (solid line) and experimental
LaFeAsOF PES spectrum (circles) from Ref.~\cite{PES}.}
\label {pd-spec} 
\end {figure}

Comparison of the LDA+DMFT single-particle spectral functions 
to various experimental spectra is presented in Figs.~\ref{d-spec}-\ref{pd-spec}.
Taking into account the selection rules for XES (X-ray emission spectroscopy) 
(neglecting the energy dependence of matrix elements) we compare the Fe  L$_3$ XES spectrum of Ref.~\cite{X-ray}, 
which corresponds to $2p \rightarrow 3d$ transitions, with the calculated LDA+DMFT 
Fe-$3d$ spectral function (see Fig.~\ref {d-spec}) to find a good agreement between the two.  
The shoulder in the experimental curve
near -2.5 eV corresponds to the low energy peak in the calculated spectrum originating from
strong hybridization between the Fe-$d$ and As-$p$ states (see also Fig.~\ref
{fig1}). 
In Fig.~\ref {pd-spec} we present the total LDA+DMFT spectral function together with the experimental photoemission 
data of Ref~\cite{PES}. Again we find a very good agreement between the theory and experiment. 
The sharp peak at the Fermi energy corresponds to a partially filled Fe-$d$ band while the broad feature 
between -2 and 6 eV corresponds to the oxygen and arsenic $p$ bands.

In conclusion, we have calculated the average Coulomb interaction $U$ and $J$ within
the Fe $d$-shell in LaFeAsO using the constrained
DFT procedure in the basis of Wannier functions and have obtained the Coulomb parameters values $F^0$=3.5~eV, $J$=0.8~eV . 
The LDA+DMFT calculations  yield moderately correlated iron $d$ bands in this compound.
This conclusion is supported by spectroscopic studies
of this material and other pnictides.

Support by the Russian Foundation for Basic Research under Grant No.
RFFI-07-02-00041, President of Russian Federation fund of support for scientific schools (grants NSH-1929.2008.2 and NSH-1941.2008.2), the Natural Sciences and Engineering Research Council of Canada (NSERC), and the Canada Research Chair program is gratefully acknowledged. 


\end {document}